\documentclass[preprint2]{aastex}

\newcommand{\kms}{kms$^{-1}$}

\shorttitle{High-Latitude CO}
\shortauthors{Lee et al.}

\begin{document}


\title{Discovery of High-Latitude CO in a HI Supershell in NGC~5775}


\author{S.-W. Lee and E. R. Seaquist}
\affil{Department of Astronomy and Astrophysics, University of Toronto,
    Toronto, Ontario M5S 3H8, Canada}
\email{swlee@astro.utoronto.ca, seaquist@astro.utoronto.ca}

\author{S. Leon\altaffilmark{1}}
\affil{I. Physikalisches Institut, University of Cologne, 
Germany}
\email{sleon@ph1.uni-koeln.de}

\author{S. Garc\'{\i}a-Burillo}
\affil{Observatorio Astron\'{o}mico Nacional-OAN, Apartado 1143, 28800
Alcal\'{a} de Henares-Madrid, Spain}
\email{burillo@oan.es}

\and 

\author{J. A. Irwin}
\affil{Department of Physics, Queen's University, Kingston, Ontario
K7L 3N6, Canada}
\email{irwin@astro.queensu.ca}

\altaffiltext{1}{Present Address: Instituto de Astrofmsica de
Andalucma (CSIC), c/ Camino Bajo de Huitor, 24, Apartado 3004, 
18080 Granada; email: stephane@iaa.es}




\begin{abstract}
We report the discovery of very high latitude molecular gas in the
edge-on spiral galaxy, NGC~5775. Emission from both the J=1-0 and 2-1
lines of $^{12}$CO is detected up to 4.8~kpc away from the mid-plane
of the galaxy. NGC~5775 is known to host a number of HI
supershells. The association of the molecular gas (M$_{H_2,F2}$ =
3.1$\times$10$^7$~M$_{\sun}$) reported here with one of the HI
supershells (labeled F2) is clear, which suggests that molecular gas
may have survived the process which originally formed the
supershell. Alternatively, part of the gas could have been formed in
situ at high latitude from shock-compression of pre-existing HI
gas. The CO J=2-1/J=1-0 line ratio of 0.34$\pm$40\% is significantly
lower than unity, which suggests that the gas is excited subthermally,
with gas density a few $\times$ 10$^2$~cm$^{-3}$. The molecular gas is
likely in the form of cloudlets which are confined by magnetic and
cosmic rays pressure. The potential energy of the gas at high latitude
is found to be 2$\times$ 10$^{56}$~ergs and the total (HI + H$_2$)
kinetic energy is 9$\times$ 10$^{53}$~ergs. Based on the energetics of
the supershell, we suggest that most of the energy in the supershell
is in the form of potential energy and that the supershell is on the
verge of falling and returning the gas to the disk of the galaxy.
\end{abstract}


\keywords{galaxies: halos --- galaxies: ISM --- 
galaxies: individual (NGC 5775) --- galaxies: spiral --- ISM: bubbles }

\section{Introduction}
Galactic HI shells and supershells, distinguished by whether their
initial energy requirement is less or more than 10$^{53}$~ergs, were
first studied by \citet{hei79,hei84}. The large energies found in the
supershells ($>$ 10$^{53}$~ergs) imply that these structures must have
a tremendous influence on the structure of the interstellar medium. In
addition, supershells that break through the gaseous disk to reach
high galactic latitudes may be a source of star formation in the
halo. For example, supershells may act as ''chimneys'' through which
hot gas from the disk funnels to the halo \citep[e.g., the ''chimney
model'',][]{nor89}. This hot gas may cool and eventually form stars in
the halo.  Alternatively, the molecular gas in the supershell may
reach high-latitude and directly provide raw material for star
formation in the halo. Although which of these two scenarios is at
work cannot be distinguished easily, the study of molecular gas in
supershells still provides an important clue to high-latitude star
formation. Molecular gas at high latitude also presents an important
aspect in the understanding of the global evolution of the
interstellar medium in spiral galaxies. In the Milky Way Galaxy, the
study of supershells is hindered by difficulties with distance
determination and the resulting confusion. In external galaxies, these
problems are minimized.

\begin{figure}
\plotone{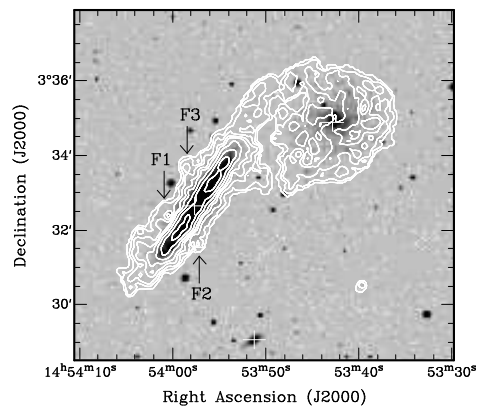}
\caption{HI column density contours superimposed on the Digital Sky
Survey image of the vicinity of NGC~5775. The white crosses indicate
the centres of NGC~5775 (left centre), NGC~5774 (upper right) and
IC~1070 (bottom). Contour levels are at 1, 5, 10, 17.5, 25, 40, 60,
and 82.6$\times$10$^{20}$~cm$^{-2}$. The HI supershells are labeled
F1 to F3. \label{hi}}
\end{figure}

NGC~5775 is an edge-on ($i$ = 86\arcdeg), infrared-bright (L$_{FIR}$ =
2.6$\times$10$^{10}$~L$_{\sun}$) galaxy at a distance of
24.8~Mpc. \citet{irw94} observed this galaxy and its face-on neighbour
NGC~5774 in HI using the VLA and provided models for their HI
distributions. She showed that the two galaxies may be engaging in an
early phase of interaction, with two HI bridges connecting them.
Numerous HI arcs and extensions beyond the disk of NGC~5775 are also
observed. Six HI supershells were cataloged by \citet{lee98}. In a
multi-wavelength study of NGC~5775, \citet{lee01} report spatial
correlations of HI, radio continuum, X-rays and far-infrared emission
at the positions of the three largest HI supershells, labeled F1
through F3 in Fig.~\ref{hi}.

\section{Observations}
Observations were performed with the IRAM 30m telescope \footnote{The
Institut de RadioAstronomie Millim\'{e}trique (IRAM) is cofunded by
the Centre National de la Recherche Scientifique, France (CNRS), the
Max Planck Gesellschaft, Germany (MPG), and the Instituto
Geogr\'{a}fico Nacional, Spain (IGN).} on 2000 October, 2001 April and
November. The CO J=1-0 line at 115~GHz and J=2-1 line at 230~GHz
were observed simultaneously using two dual mixers tuned in single
sideband mode. The observations were done in ``wobbler'' switching
mode with a wobbler throw of 240\arcsec, resulting in very flat
baselines. Relevant observing and spectral parameters are listed in
Table~\ref{tab1}. Data reduction was carried out using the GILDAS
software package. Gaussian fits were obtained in order to find the
peak antenna temperatures, central velocities,
full-width-half-maximums (FWHMs) and the integrated intensities of the
lines. For clarity of presentation, we define the x and y directions
to be parallel and perpendicular to the galaxy's major axis,
respectively.  All offsets are given in the form (x-offset,y-offset)
in units of arcseconds, where the offsets are measured with respect to
the observed centre at $\alpha$(J2000) = 14$^h$53$^m$57\fs7,
$\delta$(J2000) = 3\arcdeg32\arcmin40\farcs0. Note that x-offset$>$0
represents the eastward direction and y-offset$>$0 represents the
northward direction.

\begin{table}
\begin{center}
\caption{Observing \& Spectra Parameters. \label{tab1}}
\begin{tabular}{lcc}
\tableline\tableline
Parameter  & 115~GHz & 230~GHz \\
\tableline
T$_{sys}$\tablenotemark{a} (K) & 250 & 350 \\
HPBW\tablenotemark{b} (\arcsec) & 21 & 11 \\
B$_{eff}$\tablenotemark{c} & 0.75 & 0.52 \\
F$_{eff}$\tablenotemark{d} & 0.95 & 0.91 \\
Channel Width (\kms) & 26 & 21 \\
Spectra rms (mK) & 2 & 3 \\
\tableline
\end{tabular}
\tablenotetext{a}{Typical system temperature.}
\tablenotetext{b}{Half-power-beamwidth.}
\tablenotetext{c}{Main beam efficiency.}
\tablenotetext{d}{Forward efficiency.}
\end{center}
\end{table}

\section{Results}
\label{results}
\begin{figure}
\plottwo{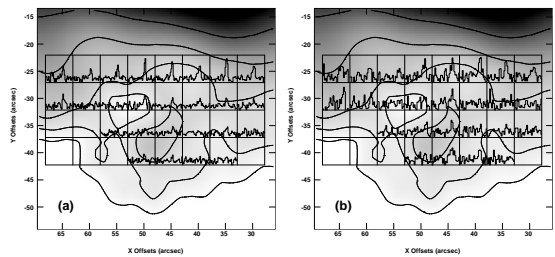}{figure2b.ps}
\caption{(a) Observed CO J=1-0 spectra superimposed on the HI column
density map of F2. The HI contour between x-offset = 50\arcsec\ and
55\arcsec\ and y-offset = -35\arcsec\ and -30\arcsec\ represents a
decrease in column density. Horizontal axes of the spectra are from
-500~\kms\ to 500~\kms\ and vertical axes are plotted from -7~mK to
40~mK in the T$_{MB}$ temperature scale. (b) As in (a) but for CO
J=2-1. Horizontal axes of the spectra are plotted from -300~\kms\ to
300~\kms\ and the vertical axes from -7~mK to 28~mK.
\label{cospectra}}
\end{figure}

We initially searched for CO emission at a few selected positions in
all three HI supershells (see Fig.~\ref{hi}). Although all three
showed tentative detections, the spectral lines at F2 are the most
obvious. We therefore proceed to map the CO emission in F2. The
observed $^{12}$CO J=1-0 and J=2-1 spectra are presented in
Fig.~\ref{cospectra}, superimposed on the HI total intensity
map. Emission from both CO transitions is detected up to 40$\arcsec$
(4.8~kpc) away from the mid-plane of the galaxy. Within the region
(35,-25) to (60,-40), roughly the size of the CO J=1-0 beam, the
average peak values of T$_{MB}$ for CO J = 1-0 and 2-1 are about 13~mK
(3.2 sigma) and 13.5~mK (2.3 sigma), respectively. The fact that there
is an absence of CO emission near the HI ``hole'' at the offset of
(55,-30) (see Fig.~\ref{cospectra}) suggests that the molecular gas
mimics the shell-like distribution of the HI feature F2.

Emission from a flared or warped disk may resemble gas at
high-latitude when seen in projection. However, in such a case the
line-of-sight central velocity of the emission line, which presumably
originates from the outer part of the disk, would be much closer to
the systemic velocity of the galaxy. We therefore compare the average
central velocity and average FWHM of all CO 1-0 spectra in F2 to a
spectrum along the major axis. The average central velocity and
average FWHM at F2 are 131$\pm$19~\kms\ and 83$\pm$23~\kms,
respectively. The uncertainties are the standard deviations of all the
spectra and are both smaller than the velocity resolution. At (40,0)
(spectrum not shown), the corresponding values are 126~\kms\ and
69~\kms, respectively. The central velocities comparison shows that
the gas in F2 shares the same circular velocity as the major axis gas
at the same galactocentric distance and does not originate from the
outer disk. The narrow line-widths support this, since gas from the
outer disk would have wider spectra. Therefore, we conclude that the
gas associated with F2 is indeed at high galactic latitude and not
from a flared or warped disk.  Detailed analysis of the kinematic
structures of the galaxy and the supershell will be forthcoming.

In order to study the molecular gas that truly belongs to the
supershell at high latitude, we have to eliminate the contribution of
CO from the disk. We assume that the vertical CO distribution of the
galaxy can be represented by a single gaussian which peaks at the
mid-plane, and that the emission from the supershell at high latitude
is superimposed on the wing of the gaussian. The emission from the
supershell can then be isolated by subtracting the gaussian model from
the observed data. To fit the gaussian, we average the CO integrated
intensity map from x-offsets = 30\arcsec\ to 65\arcsec, to increase
the signal-to-noise. A gaussian was fitted from y-offset = +20\arcsec\
to -20\arcsec\ (hence does not include the high-latitude
emission). The results of the fits are good, the residuals being less
than 5\% of the peak values in both transitions. At the centre of the
supershell (y-offset = -32.5\arcsec) , the difference in integrated
intensities between the data and the model gives the CO emission from
the supershell alone. The total (disk contribution plus supershell
contribution) integrated intensities within a 21\arcsec\ beam at the
centre of F2 are 1.28 and 1.05~K-\kms\ for the 1-0 and 2-1
transitions, respectively, where the temperatures are in the T$_{MB}$
scale. The uncertainties in these values, taking into account noise,
calibration and baseline errors, are about 30\%. For proper
comparison, the CO J = 2-1 data were smoothed to the resolution of the
1-0 data. After subtracting the disk contribution, the CO 1-0 and 2-1
integrated intensities are given by I$_{10}$ = 0.89~K-\kms\ and
I$_{21}$ = 0.30~K-\kms, respectively. The line ratio of the two
transitions, I$_{21}$/I$_{10}$, is therefore 0.34$\pm$40\%.

For comparison, we obtained the line ratio at mid-plane corresponding
to the same galactocentric radius of F2. We use our JCMT CO J = 2-1
integrated intensity (= 17.4$\pm$20\%~K-\kms) at the position equivalent to
(40,0) \citep{lee98}, and the IRAM 30m CO J = 1-0 integrated intensity
(= 20.2$\pm$20\%~K-\kms) at the same position. The JCMT beam has the same
HPBW as the IRAM 30m beam at these frequencies. The resultant line
ratio at the mid-plane is 0.86$\pm$30\%.

The total molecular gas mass in F2 (excluding the disk contribution),
obtained using the CO J=1-0 integrated intensity, is found to be
M$_{H_2,F2}$ = 3.1$\times$10$^7$~M$_{\sun}$ within the 21\arcsec\ beam
($\equiv$ 2.5~kpc in diameter), assuming a CO-H$_2$ conversion factor
of 3$\times$10$^{20}$~cm$^{-2}\cdot$K-kms$^{-1}$ \citep{you91}. Since
the CO and the HI distributions within F2 agree fairly well, we have
probably detected most of the CO emission in the supershell (see
Fig.~\ref{cospectra}).

We have also obtained the HI integrated intensity within a
21\arcsec-beam at F2 following the same procedure outlined above. The
HI data have been published in \citet{irw94} and in \citet{lee01}. The
HI integrated intensity, after subtracting the disk contribution, is
0.11~Jy/beam$\cdot$\kms\ and the corresponding HI mass is M$_{HI,F2}$
= 1.6$\times$10$^7$~M$_{\sun}$, or 50\% of M$_{H_2,F2}$. The total gas
mass (HI + H$_2$) in F2 is therefore 4.7$\times$10$^7$~M$_{\sun}$.

\section{Discussion}
This paper reports the highest latitude (4.8~kpc) molecular gas
detected to date in any galaxy. In our own Galaxy, high-latitude
molecular clouds are found in association with the HI shells around
two OB associations, the Per OB3 and the Sco OB2 associations
\citep{bha00}; and in association with a hot bubble between Cepheus
and Cassiopeia \citep{gre89}. There are only a few examples of
detections of {\it discrete} extraplanar molecular gas in external
galaxies. In NGC~891, a molecular spur extending to 0.5~kpc above the
mid-plane was detected by \citet{han92}. Possible formation mechanisms
considered for this structure include ejection by a superbubble or the
buoyancy force on the gas due to Parker instability. Recently,
molecular gas up to 3~kpc above the mid-plane was discovered in the
molecular outflow in M~82 (CO detection by \citet{tay01}; SiO
detection by \citet{gar01}).  Many of these examples of discrete
high-latitude molecular gas are thought to be related to star
formation activities in the disk such as stellar wind or supernovae
swept-up shells. At the highest latitudes, the molecular gas may
represent the component of the shells that broke through the thin disk
to reach the halo.

\subsection{The Physical Conditions in the Supershell}
The line ratio (0.34$\pm$40\%) for the extraplanar gas obtained is
significantly lower than the corresponding ratio at the
mid-plane. Even taking into account the uncertainty, its highest
possible value is less than 0.5. Assuming that the gas is thermalized
so that the excitation temperature of all rotational levels is the
same, the interpretation of such a low line ratio is that the
excitation temperature is low (T$_{ex} \sim$ 5~K). Although no estimate
of the gas or dust temperature in the supershell is currently known,
the possibility of shocks, deduced from the trend of radio continuum
spectral indices \citep[see][]{lee01}, suggests that the gas kinetic
temperature may be $>$ 5~K due to shock heating. In this case, the gas
density must be low ($\sim$ a few hundred particles per cm$^{3}$) so
that thermalization does not occur. Using a Large Velocity Gradient
(LVG) analysis, we find that such a low line ratio is consistent with
a gas density n$_{H_2} < 1000$~cm$^{-3}$, although we cannot constrain
the gas kinetic temperature. It is noted that the line ratio is
expected to be closer to one in shock regions due to the elevated gas
kinetic temperature, which tends to populate the upper rotational
levels. Better constraints of the physical parameters will have to
await data of higher CO transitions.

Given the gas density and gas mass within the 21\arcsec\ beam, the
filling factor of the molecular gas is of order 10$^{-3}$. That is,
the molecular gas must be in the form of small cloudlets. We can
estimate whether these cloudlets are in pressure equilibrium with the
surrounding environment. The ambient pressure surrounding the
cloudlets comes from the magnetic pressure (P$_m$), cosmic rays
pressure (P$_{CR}$), and the gas pressure. The first two are equal in
the case of energy equipartition and are both given by B$^2$/8$\pi$,
where B $\sim$ 3~$\mu$G is the equipartition magnetic field
\citep{dur98}. The gas pressure is the combined ionized (P$_i$) and
neutral gas pressure (P$_{HI}$), each given by P = nkT, where n is the
gas number density, k is the Boltzmann constant and T is the kinetic
temperature of the gas.  We assume the ionized gas is at T = 10$^4$~K
and the electron density n = 9$\times$10$^{-4}$~cm$^{-3}$ at 4~kpc
above midplane (from the electron density distribution given in
\citep{col00}). For the HI gas we use T = 8000~K and density n$_{HI}$
= 2$\times$10$^{-3}$~cm$^{-3}$ at 4 ~kpc above midplane (from the
vertical density distribution given in \citet{irw94}). From these
values, we obtain P$_m$ $\sim$ P$_{CR}$ $\sim$
4$\times$10$^{-13}$~erg~cm$^{-3}$, P$_i$ =
2$\times$10$^{-15}$~erg~cm$^{-3}$, and P$_{HI}$ =
2$\times$10$^{-15}$~erg~cm$^{-3}$. The averaged internal pressure of
the molecular gas is P$_{cloud}$ = 1$\times$10$^{-12}$~erg~cm$^{-3}$,
using parameters given in the previous paragraph. It seems that the
magnetic and cosmic rays pressure combined is sufficient to confine
the clouds without internal gravitation.

\subsection{Origin of Molecular Gas at High Latitude}
Observations of molecular gas at high-latitude are interesting in that
they may explain high-latitude star formation. The key question is:
how does the molecular gas reach such high latitude? Two possibilities
may be considered. First, the molecular gas was formed in the disk and
was ejected by the outflow that formed the HI supershell; and second,
the molecular gas was formed at high-latitude. Of course, a
combination of these is also possible.

The gravitational potential energy (E$_{PE}$) of a cloud at a height,
z, above the galactic plane is calculated by solving the Poisson
equation for a disk flattened in one-dimension. We have assumed the
galactic value for the stellar scale height and the solar
neighbourhood value for the stellar mass density at mid-plane (325~pc
and 0.175~M$_\sun$pc$^{-3}$, respectively, from \citet{fre87}) since
the corresponding values for NGC~5775 are not known.  Including both
HI and H$_2$ gas, E$_{PE}$ = 2$\times$10$^{56}$~ergs in F2.  Note that
E$_{PE}$ scales as the square of the stellar scale height so that if
the scale height of NGC~5775 is twice that of the Milky Way, E$_{PE}$
will increase by a factor of 4. We estimate the total kinetic energy
(E$_{KE}$) of the gas (HI + H$_2$) to be E$_{KE}$ =
9$\times$10$^{53}$~ergs. This includes the kinetic energy from the HI
gas in F2, which is found to be 6$\times$10$^{53}$~ergs, using
M$_{HI,F2}$ calculated above and assuming the expansion velocity is
62.5~kms$^{-1}$ \citep{lee01}; and the kinetic energy from the H$_2$
gas, which is found to be 4$\times$10$^{53}$~ergs. The expansion
velocity of the H$_2$ gas is taken to be half the averaged FWHM of all
the CO J=1-0 spectra (see \S~\ref{results}) in F2 minus, in
quadrature, the expected linewidth from differential rotation within
the beam (44~\kms) and the channel width (26~\kms), and is equal to
32.5~\kms. The predicted differential rotation within the beam is
calculated using the HI rotation curve model given in
\citet{irw94}. Assuming that the supershell expands isotropically (a
fair assumption based on the near circular morphology of F2), then the
line-of-sight expansion velocity observed is a good estimate of the
true expansion velocity of the supershell. Therefore, E$_{KE}$ of the
supershell is more than two orders of magnitude lower than
E$_{PE}$. The large difference in E$_{PE}$ and E$_{KE}$ suggests that
the supershell is being observed at a time when it is about to fall
back to the disk so that most of the energy is in the form of
potential energy. We can therefore visualize an explosion at the
mid-plane that created an expanding bubble of hot gas. In the process,
neutral gas (atomic and molecular) is entrained in the ejecta, forming
the supershell, reaching large z-height. Under the influence of the
gravitational potential, the supershell is now about to plunge back,
returning the gas to the star-forming disk. Such a recycling process
has been proposed in the galactic fountain model \citep{bre80}.

Another possibility for the molecular gas to exist at high latitude is
that H$_2$ gas can be formed in situ via shock-compression
\citep[e.g.,][]{mag87,elm88}, occurring within HI pre-existing in this
region. The effect of a strong shock propagating through the
interstellar medium is that dense molecular cloudlets can form due to
thermal instability in the shock regions \citep{koy00}. The existence
of H$\alpha$ emission \citep[possibly due to shock-ionization,
see][]{lee01} in F2 suggests that at least some of the molecular gas
may be formed this way. It is more difficult to explain how the HI
come to reside at high-latitude, in view of the large mechanical
energy required (equals to E$_{PE}$). Various mechanisms have been
examined in \citet{lee01}, including supernova explosions and cloud
impacts. In any case, this material would likely have originated
within NGC 5775 since the mean radial velocity agrees with the orbital
velocity of the disk. Finally, assuming typical Galactic dust-to-gas
ratio, the high-latitude molecular clouds could be emitting
substantial infrared radiation and has indeed been observed at
850$\mu$m (Brar et al., private communication).

\section{Conclusions}
This paper reports the detection of high-latitude molecular gas in
NGC~5775. The shell-like distribution of the CO emission coincides
exactly with that of the HI supershells, suggesting that we have
detected the molecular shell associated with the HI supershell. The
existence of the molecular shell means that molecules are not
destroyed during the ejection of the supershell and is entrained in
the expanding flow to reach high latitude. Some of the molecular gas
may have been formed in situ, via shock-compression of pre-existing HI
gas. The CO J=2-1/1-0 line ratio (0.34$\pm$40\%) suggests that the gas
density in the supershell is low and the gas is subthermally
excited. The molecular gas is probably in the form of cloudlets which
are confined by magnetic and cosmic rays pressure. Based on energetics
ground, we proposed that the supershell may be at a stage where it is
about to plunge towards the disk of the galaxy, returning the gas to
the bulk of the gas reservoir of the galaxy.

\acknowledgments
This work was supported in part (to S.L.) by the Deutsche
Forschungsgemeinschaft (DFG) via grant SFB 494, by special funding
from the Science Ministry of the Land Nordrhein-Westfalen. It is also
supported by a grant (to E.R.S.) from the National Sciences and
Engineering Research Council of Canada.

\end{document}